\documentclass[11pt]{article}

\usepackage[]{ACL2023}
\usepackage{times}
\usepackage{latexsym}

\usepackage[T1]{fontenc}

\usepackage[utf8]{inputenc}

\usepackage{microtype}
\usepackage{pdflscape}
\usepackage{rotating}
\usepackage{amssymb}
\usepackage{inconsolata}

\AtBeginDocument{%
  }

\usepackage{amsmath,amsfonts}
\usepackage{graphicx}
\usepackage{textcomp}
\usepackage{xcolor}
\usepackage{caption}
\usepackage{lipsum}
\usepackage{makecell}
\usepackage{xspace}
\usepackage{float}
\usepackage{tipa}

\usepackage{arydshln}
\usepackage{enumitem}
\usepackage{multirow}
\usepackage{tcolorbox}
\usepackage{colortbl}
\usepackage{adjustbox} 
\usepackage[T1]{fontenc}

\usepackage{threeparttable}
\usepackage{listings}
\usepackage{xcolor}
\usepackage{booktabs}
\usepackage{subcaption}
\usepackage{algorithm}
\usepackage{algorithmic}

\usepackage{changes}
\usepackage{tabularx}
\usepackage{amssymb} 
\usepackage{bbding}
\usepackage{enumitem}

\newcommand{\taskname}{MRWeb\xspace}
\definecolor{lightblue}{rgb}{0.2, 0.2, 0.8}
\definecolor{lightgreen}{rgb}{0.2, 0.8, 0.2}

\newsavebox{\arrangebox}

\begin{document}

\title{MRWeb: An Exploration of Generating Multi-Page Resource-Aware Web Code from UI Designs}

\author{
    {\bf Yuxuan Wan}$^{1}$,
    {\bf Yi Dong}$^{1}$,
    {\bf Jingyu Xiao}$^{1}$, 
    {\bf Yintong Huo}$^{1}$, \\
    {\bf Wenxuan Wang}$^{1}$\thanks{\ \ Wenxuan Wang is the corresponding author.},
    {\bf Michael R. Lyu}$^{1}$ \\
    $^1$The Chinese University of Hong Kong, Hong Kong, China \\
    \texttt{\{yxwan9, ythuo, wxwang, lyu\}@cse.cuhk.edu.hk},\\ \texttt{dy13367795706@gmail.com, whalexiao99@gmail.com}
}

\maketitle

\begin{abstract}
Multi-page websites dominate modern web development. However, existing design-to-code methods rely on simplified assumptions, limiting to single-page, self-contained webpages without external resource connection. To address this gap, we introduce the Multi-Page Resource-Aware Webpage (MRWeb) generation task, which transforms UI designs into multi-page, functional web UIs with internal/external navigation, image loading, and backend routing. We propose a novel resource list data structure to track resources, links, and design components. Our study applies existing methods to the MRWeb problem using a newly curated dataset of 500 websites (300 synthetic, 200 real-world).
Specifically, we identify the best metric to evaluate the similarity of the web UI, assess the impact of the resource list on MRWeb generation, analyze MLLM limitations, and evaluate the effectiveness of the MRWeb tool in real-world workflows. The results show that resource lists boost navigation functionality from 0\% to 66\%-80\% while facilitating visual similarity. Our proposed metrics and evaluation framework provide new insights into MLLM performance on MRWeb tasks. We release the MRWeb tool, dataset, and evaluation framework to promote further research\footnote{\url{https://github.com/WebPAI/MRWeb}}.

\end{abstract}

\section{Introduction}

\begin{figure}
    \centering
    \includegraphics[width=\linewidth]{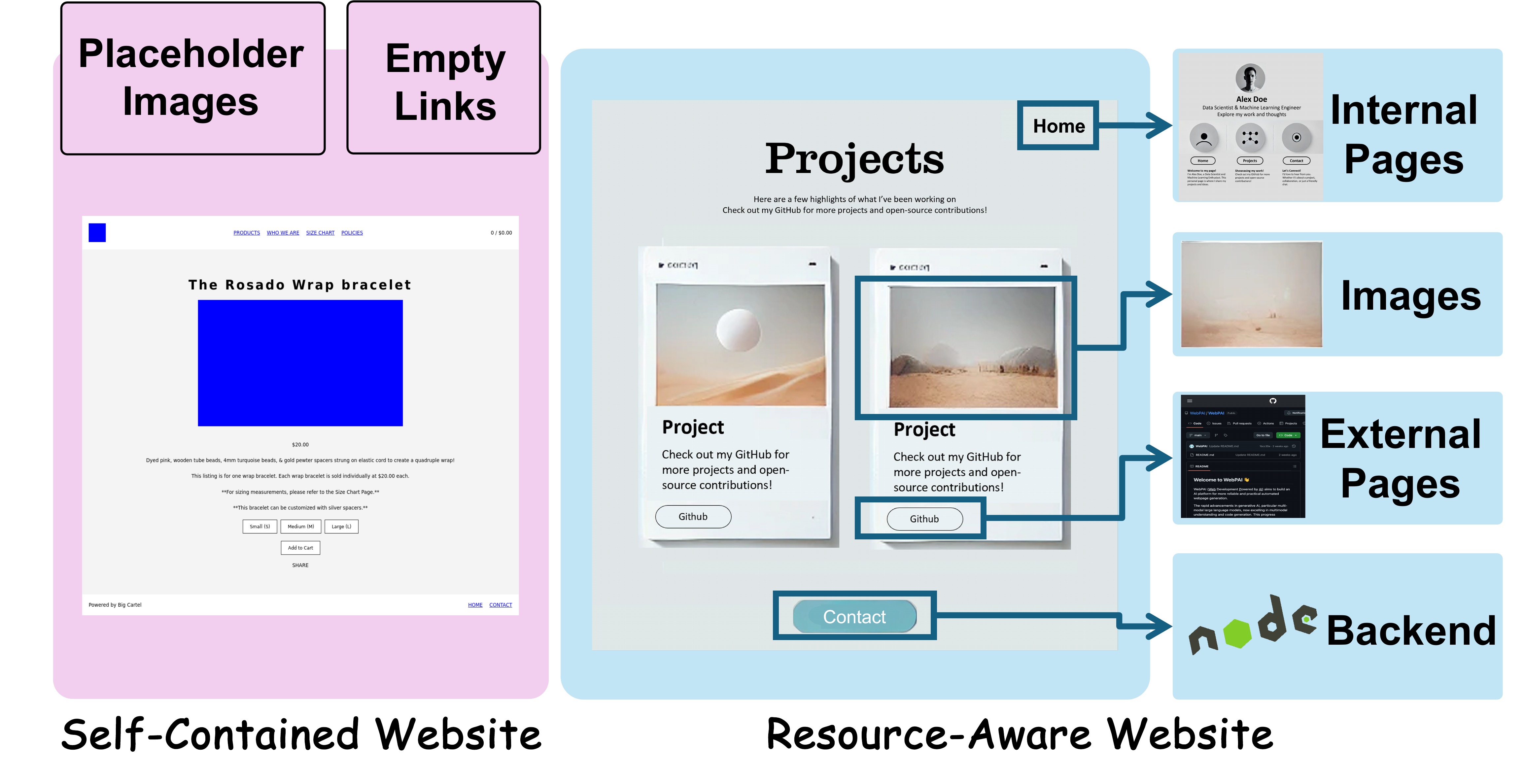}
    \vspace{-10pt}
    \caption{Comparison between self-contained webpage and multi-page resource-aware webpage (MRWeb). MRWeb supports multi-page navigation, real-image loading and backend routing.}
    \label{fig:pipeline}
    \vspace{-15pt}
\end{figure}

Websites are important in today's digital landscape, serving as essential platforms for diverse applications in our daily lives~\cite{website_statistics_2024}.

They can be categorized into two types: single-page and multi-page. Single-page websites update content dynamically on a single page without reloading, while multi-page websites consist of multiple interconnected pages. In reality, multi-page websites are dominant due to their scalability and structured navigation functionality~\cite{waytogrow2024spa, web2024onepage}. To examine this, we conducted a preliminary study by sampling the top 300 most visited websites ranked by the Tranco list~\footnote{https://tranco-list.eu}. For each website, we analyzed its structure by recursively visiting all internal links to identify distinct internal pages, as well as counting the number of external links, internal links, and images present on each webpage. The results showed that 271 (90.3\%) of the sampled websites are multi-page, highlighting their widespread usage and the complex structures they often exhibit. Appendix~\ref{sec:motivating-study} describes the study in detail.

Practical solutions for generating multi-page web user interfaces (UIs) from designs remain underexplored. Existing work focuses on the simpler task of design-to-code, which generates self-contained web pages (Figure~\ref{fig:pipeline}). This leaves a gap in addressing the complexities of multi-page, resource-aware web UI development, where \textbf{(1) webpages link to internal pages with unlimited navigation paths, (2) webpages link to external resources like websites and images, and (3) webpages route to backends for data exchange}. We term this task \underline{\taskname} to highlight the shift from design-to-code to \underline{M}ulti-Page \underline{R}esource-Aware \underline{Web}page generation. Figure~\ref{fig:pipeline} contrasts self-contained webpages with MRWebs, which \textbf{enable code-free development from UI designs to resource-aware, navigable websites, democratizing web development}.


Unfortunately, constructing such a framework poses several challenges. First, there is no established data structure that integrates visual design elements with internal or external resources and tracks their correspondences. Unlike the design-to-code task that generates self-contained code from design images, \taskname should incorporate resources, navigation paths, and their links to visual design elements, necessitating a new data structure. 
Second, there is a lack of high-quality datasets for \taskname since the previousn-to-code task did not contain real images or internal/external links. 
Third, there are no standardized metrics for evaluating the performance of the \taskname task.  This makes it difficult to measure the accuracy of generated links, images, and their correspondence with visual design elements. As a result, there is insufficient understanding of how effectively MLLMs produce \taskname code from designs.

To bridge this gap, we propose the first evaluation framework for \taskname generation. Specifically, we define a novel data structure, the \textbf{resource list}, which uses a dictionary-like format to store internal/external resources, such as links and images, and their correspondence with visual designs (e.g., screenshots). Next, we collect the first \taskname dataset, comprising 500 websites (300 synthetic, 200 real-world), along with their associated resource lists, screenshots, and ground truth code. The generation framework accepts the resource lists and screenshots as inputs and calls MLLMs to generate functional \taskname code directly. To assess the generated code, we introduce a suite of metrics designed to evaluate and analyze both the visual and functional performance of MLLMs on the \taskname task. Additionally, we implement this framework as a user-friendly tool for \taskname generation, releasing all code and data to encourage future research.

\begin{table*}[ht]
    \centering
    \caption{Comparison and statistics of benchmarks and datasets. All statistics are in the ``average $\pm$ standard deviation'' format.}
    \label{tab:comparison}
    \vspace{-5pt}
    \begin{adjustbox}{width=\textwidth}
    \begin{tabular}{|l|c|c|c|c|c|c|}
        \toprule
        & \textbf{WebSight} & \textbf{VISION2UI} & \textbf{Design2Code} & \textbf{DWCG} & \textbf{Interaction2Code} & \textbf{MRWeb } \\
        & \cite{laurençon2024unlocking} & \cite{Gui2024VISION2UIAR} & \cite{Si2024Design2CodeHF} & \cite{Yun2024Web2CodeAL} & \cite{xiao2024interaction2code} & (Ours) \\
        \midrule
        \textbf{Scope} & Self-contained & Self-contained & Self-contained & Self-contained & Self-contained & \taskname \\
        \textbf{Int. navigation} & \XSolidBrush & \XSolidBrush & \XSolidBrush & \XSolidBrush& \XSolidBrush & \Checkmark \\
        \textbf{Ext. navigation } & \XSolidBrush & \XSolidBrush & \XSolidBrush & \XSolidBrush & \XSolidBrush & \Checkmark \\
        \textbf{Backend routing} & \XSolidBrush & \XSolidBrush & \XSolidBrush & \XSolidBrush & \XSolidBrush & \Checkmark \\
        \textbf{Real img insertion} & \XSolidBrush & \XSolidBrush & \XSolidBrush & \XSolidBrush & \XSolidBrush & \Checkmark \\
        \midrule
        \textbf{Source} & Synthetic & Real-World & Real-World & Synthetic & Real-World & Synthetic / Real-World \\
        \textbf{Size} & 823K & 20k & 484 & 60K & 97 & 500 (300 / 200) \\
        \textbf{Avg. Len (tokens)} & $647\pm216$ & $8460\pm7120$ & $31216\pm23902$ & $471.8\pm162.3$ & $141084\pm160438$ & $692\pm227$ / $113724\pm139761$ \\
        \textbf{Avg. Tags} & $19\pm8$ & $175\pm94$ & $158\pm100$ & $28.1\pm10.6$ & $1291\pm1574$ & $18\pm7$ / $543\pm768$ \\
        \textbf{Avg. DOM Depth} & $5\pm1$ & $15\pm5$ & $13\pm5$ & $5.3\pm1.0$ & $18\pm6$ & $5\pm1$ / $15\pm6$ \\
        \textbf{Avg. Unique Tags} & $10\pm3$ & $21\pm5$ & $22\pm6$ & $13.6\pm2.7$ & $31\pm9$ & $10\pm3$ / $22\pm8$ \\
        \textbf{Avg. Resource List Len.} & - & - & - & - & - & $3\pm2$ / $92\pm130$ \\
        \bottomrule
    \end{tabular}
    \end{adjustbox}
\vspace{-15pt}
\end{table*}

To sum up, the contributions of this study are:
\begin{itemize}[leftmargin=*]
    \item Define the \taskname problem, introduce the first \taskname benchmark with an innovative resource list data structure, and construct the first \taskname dataset consisting of \textit{500} websites.
    \item Propose a suite of metrics, conduct the first comprehensive image quality assessment (IQA) in the web UI domain to determine the best evaluation metric for web UI code generation, and collect an annotated IQA dataset for future studies.
    \item Conduct a comprehensive study to evaluate the performance of SOTA MLLMs on \taskname generation and highlighting some challenges faced by MLLMs.
    \item Develop a user-friendly tool for \taskname tasks and release all code and data to foster further research and development in this emerging area.
\end{itemize}

\section{Related Work}
Existing UI code generation methods operate on three unrealistic assumptions: (1) websites have a single or limited number of pages, (2) webpages lack external links, and (3) webpages do not interact with backends. Our study in Appendix~\ref{sec:motivating-study} highlights the misalignment of these assumptions with real-world web UI needs. Table~\ref{tab:comparison} provides a comparison of our benchmark with existing works, contextualizing our contribution.  A detailed discussion of related works is in Appendix~\ref{appendix:related-work}.

\section{Task Formulation} 

\paragraph{Resource List} The \textit{resource list} serves as a structured representation of a webpage's navigational and visual elements, such as hyperlinks, images, and backend routing. Each entry in the resource list includes attributes like position, type, and URL, for instance, \texttt{\{position: $bonding\_box$, type: image, link:/dog.png\}}. This structure is crucial for enabling MRWeb generation to replicate navigational features and image sources accurately. Without the resource list, MLLMs would generate static replicas lacking interactivity and navigation. In our experimental setup, resource lists are extracted automatically using Python Selenium\footnote{\url{https://selenium-python.readthedocs.io/}}. For real-world applications, we developed an intuitive user interface that allows users to highlight actionable elements by drawing bounding boxes and inputting the corresponding resource information, as illustrated in Appendix~\ref{appendix:demo}.

\paragraph{Task Definition} Let the ground-truth webpage's HTML+CSS code be $C_0$, screenshot be $I_0$, and resource list be $R_0$, the \textit{MRWeb generation} task uses an MLLM $M$ to produce HTML+CSS code $C_g = M(I_0, R_0)$ that approximates $C_0$. The quality of $C_g$ is assessed by comparing the generated resource list $R_g$ with $R_0$ and the screenshot $I_g$ rendered from $C_g$ with $I_0$, ensuring both functional and visual alignment.

\section{Dataset Collection}
We collect two types of data: synthetic and real-world. Synthetic data enables controlled, diverse examples, including rare edge cases, but lacks the variety of real-world content. Real-world data captures authentic webpage diversity, supporting model robustness across HTML structures and styles. Combining both data types provides a comprehensive benchmark.

This section outlines our collection of code-screenshot pairs for synthetic and real-world data, the extraction of resource lists, and the statistics of the sampled data.

\subsection{Synthetic Data Collection} To create the synthetic UI-to-MRWeb dataset, we adopt and modify the WebSight dataset~\cite{laurençon2024unlocking}. The WebSight dataset contains 2 million HTML samples and their corresponding screenshots, covering a broad spectrum of website concepts. However, it cannot be used directly for an MRWeb dataset because 1) its websites lack valid internal or external links, making navigation to other pages impossible, and 2) images on the sites are randomly loaded via the Unsplash\footnote{\url{https://source.unsplash.com/}} API. This random loading causes visual inconsistencies, as identical code can result in different visuals, complicating benchmarking. To address these issues, we enhance the WebSight dataset through link insertion and image replacement.

\paragraph{Link insertion.} Using all website links from the C4~\cite{Raffel2019ExploringTL} validation set, we create a URL list. For each HTML document, we parse its content and iterate over all hyperlink tags, assigning a randomly chosen URL from our list to each hyperlink attribute. This modification ensures that every website includes valid external links to other sites.

\paragraph{Image replacement.} To ensure consistency and diversity in visual representation, we replace random images in the WebSight dataset with static, unique images for each webpage. Using the Unsplash API, we fetch images with specific keywords, dimensions, and properties to guarantee that the pictures remain consistent yet unique.

\subsection{Real-world Data Collection} We collect real-world data by capturing and simplifying HTML content from live websites. We first collect 500 URLs of real-world websites from the C4~\cite{Raffel2019ExploringTL} validation set as our data source. However, HTML files on the web often contain non-visible noise—such as comments, scripts, and hidden content—that makes them excessively lengthy and can exceed the token limits of most models. To create the real-world UI-to-MRWeb dataset, we develop a pipeline that collects and processes HTML code and screenshots from live websites. This pipeline ensures that each webpage captures authentic and static content while maintaining a simplified HTML structure compatible with our UI-to-MRWeb benchmark. The primary steps in this pipeline include saving HTML files, filtering HTML, simplifying HTML, and capturing screenshots, as outlined below:
\begin{enumerate}[leftmargin=*]
    \item Saving HTML files: The HTML+CSS content from each URL is saved into a single HTML file, ensuring all components are intact.
    \item Filtering HTML: We discard websites that are blank or erroneous (e.g., page not found).
    \item HTML Simplification: We simplify the HTML by removing all non-visible elements, comments, and non-functional Javascripts.
    \item Final Screenshot: A final screenshot is taken after simplification, completing the real-world data pipeline.
\end{enumerate}

\subsection{Resource List Extraction} 
Resource lists capture navigational and visual elements such as links, images, and backgrounds, structured to preserve the functionality and layout of each webpage. For each webpage:
\begin{itemize}[leftmargin=*]
    \item Links (\texttt{<a>} tags): We extract each hyperlink’s position, type, and target URL.
    \item Images and Background Images: For images, including both image tags and CSS background images, we record their position and source URL.
\end{itemize}
The resource list is automatically constructed by iterating through the webpage’s elements using Python Selenium, collecting attributes for each, and verifying their visibility and functionality. 

\subsection{Automation \& Dataset Statistics}
\begin{figure}
    \centering
    \includegraphics[width=0.8\linewidth]{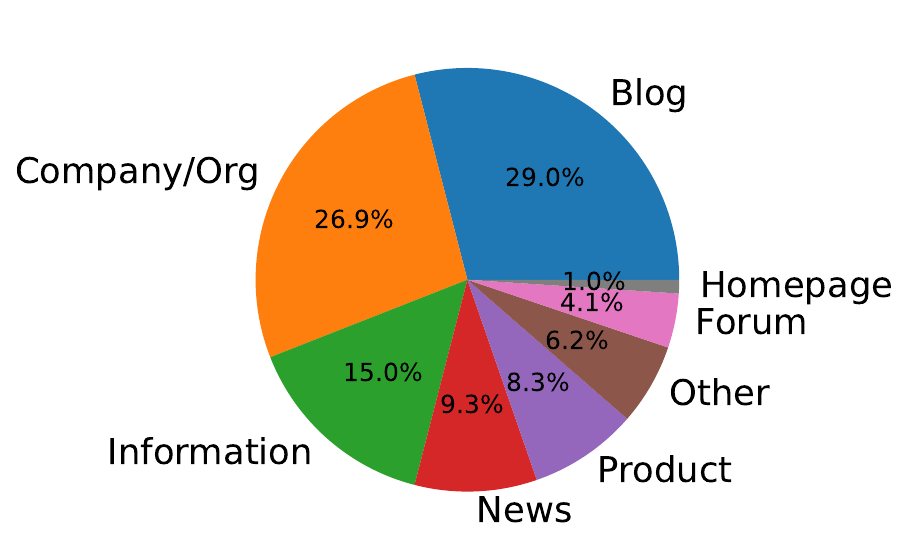}
    \caption{Topic distribution of real-world web data in \taskname dataset.}
    \label{fig:domain-distribution}
    \vspace{-15pt}
\end{figure}

We emphasize that the data collection pipeline is fully automated, enabling the on-demand generation of large-scale MRWeb training data. In principle, the synthetic dataset could match the full size of the WebSight dataset (two million), and the real-world data could encompass any website accessible on the internet. To support future research, however, we sampled 300 synthetic and 200 real-world instances. The statistics, quantitative metrics of the sampled dataset, and comparison with other datasets are provided in Table~\ref{tab:comparison}. To get a sense of the range of domains covered in our benchmark, we manually categorize what type of webpages they are based on their functions.  We present the pie chart of the most frequent domains in Figure~\ref{fig:domain-distribution}. The most prominent genres are companies' or organizations' websites and blogs.

\section{Study Setup}
\subsection{Evaluated Models}
\label{sec:setup}
We employ three state-of-the-art (SOTA) MLLMs: Gemini 1.5 \cite{google_gemini_api}, GPT-4o \cite{openai_gpt4o} and Claude-3.5 \cite{anthropic_claude} to evaluate their performance on \taskname. the specific model numbers are 20240806 for GPT-4o, 20240620 for Claude-3.5-Sonnet, and Gemini-1.5-Pro accessed during November 2024. For MLLM model configurations, we set the temperature to 0, the random seed to 42, and the $max\_tokens$ parameter to 4096 for each model. Other parameters are maintained at their default settings as specified in the corresponding API documentation~\cite{google_gemini_api_docs, openai_vision_guide, anthropic_vision_docs}.

\subsection{Prompting Strategies}
\label{subsubsec:prompt} 
We use four types of prompting methods: self-contained, zero-shot, CoT, and self-refine. Self-contained prompting is adapted from Si et al.~\cite{Si2024Design2CodeHF} to let the model directly generate code from screenshots without resource lists. This method serves as a baseline for other methods that adopt input resource lists. Zero-shot prompting directly lets the model generate HTML code from screenshots and resource lists.  Chain-of-Thought (CoT) prompting~\cite{Wei2022ChainOT} generates a chain of thought for each question and then generates the corresponding code. For CoT, we use the "let’s think step by step" instruction from Chae et al. ~\cite{Chae2024LanguageMA}. Self-refine prompting~\cite{Chen2023TeachingLL} let the model refine its own generated code via multi-turn conversation. We adopt the self-refine prompting and direct promoting method from Si et al.~\cite{Si2024Design2CodeHF}. We list the exact prompts used in our experiments in Appendix~\ref{appendix:prompts}.

\subsection{Metrics}
\label{sec:metrics}
\subsubsection{High-level Metrics}
For high-level performance, we evaluate visual similarity and functional similarity. For visual similarity, we explore three levels of image similarity metrics commonly applied in design-to-code or other computer vision (CV) tasks~\cite{Wang2004ImageQA}: pixel, structural, and semantic. The detailed background and calculation of these metrics are in Appendix~\ref{appendix:similarity}.

\noindent\textbf{Visual: Pixel-level metrics} 
\begin{itemize}[leftmargin=*]
    \item Mean Absolute Error (MAE)~\cite{Nguyen2015ReverseEM, Moran2018MachineLP}: Measures the average absolute difference in pixel intensities.
    \item Peak Signal-to-Noise Ratio (PSNR)~\cite{Lim2017EnhancedDR, Wang2019EDVRVR}: Based on Mean Squared Error (MSE), with higher values indicating greater similarity.
    \item Normalized Earth Mover's Distance (NEMD)~\cite{Arjovsky2017WassersteinG, Rubner2000TheEM}: Captures spatial differences between images, normalized to be size-independent. Higher NEMD values indicate greater similarity.
\end{itemize}

\noindent\textbf{Visual: Structure-level metrics} 
\begin{itemize}[leftmargin=*]
    \item Structural Similarity Index Measure (SSIM)~\cite{Zhou2024BridgingDA, Wang2004ImageQA}: Measures luminance, contrast, and structural changes.
\end{itemize}

\noindent\textbf{Visual: Semantic-level metrics} 
\begin{itemize}[leftmargin=*]
    \item CLIP Score~\cite{Radford2021LearningTV, Si2024Design2CodeHF}: Aligns image embeddings with language representations to capture high-level conceptual similarity.
    \item Learned Perceptual Image Patch Similarity (LPIPS)~\cite{Zhang2018TheUE, Simonyan2014VeryDC}: Assesses perceptual similarity using deep features from VGG~\cite{Simonyan2014VeryDC}.
\end{itemize}

\noindent\textbf{Functional Metric:}
\begin{itemize}[leftmargin=*]
    \item Resource Existence Ratio (RER)$_\uparrow$: The proportion of resources in the reference resource list that exist (i.e., are successfully matched to corresponding resources) in the generated list is calculated as $\text{RER} = \frac{\text{\# Matched Resources in G}}{\text{\# Total Resources in R}}$. Matching is determined based on relevant attributes of resources, such as whether navigational elements direct to the same link or whether images share the same source. 
\end{itemize}

All image pairs, except for those used with CLIP Score, are padded with random noise to ensure consistent image sizes for comparison.

\subsubsection{Fine-Grained Metrics}
Beyond assessing visual and functional similarity, we employ a suite of fine-grained metrics to evaluate the specific capabilities of MLLMs, including visual grounding, color recognition, and text extraction. For each pair of matched resources in RER, we calculate:
\begin{itemize}[leftmargin=*]
    \item Position offset$_\downarrow$: Measures the position shift between the generated and the original element with respect to the size of the original web page. For each pair of matched resources $(r_p, g_q)$, the positional alignment is evaluated by comparing the normalized offset of their corresponding web elements' center points: $\text{Position Offset} = \max\left(\frac{|x_p - x_q|}{W}, \frac{|y_p - y_q|}{H}\right)$. $(x_p, y_p)$ and $(x_q, y_q)$ are the center positions of the bounding boxes enclosing the elements; $W$ and $H$ represent the width and height of the original webpage.  

    \item Area Difference$_\downarrow$: Measures the differences in size between corresponding actionable elements with respect to the original area of the element: $\text{Area Difference} = \frac{|A_p - A_q|}{A_p}$, where $A_p$ and $A_q$ are the areas occupied by the reference and generated actionable elements. 
    
    \item Color Difference$_\downarrow$: We use the CIEDE2000 color difference formula~\cite{Luo2001TheDO} to assess the perceptual difference between the colors of element $r_p$ and $g_q$.

    \item Text Difference$_\downarrow$: For resources that involve text, such as buttons, their text similarity $\text{\textit{Text Sim}}(r_p, g_q)$ is calculated by normalizing the number of matching characters by the total length of the text. We calculate the text difference by 1 - $\text{\textit{Text Sim}}(r_p, g_q)$.
\end{itemize}
\begin{table*}[ht]
\centering
\footnotesize
\caption{Overall correlation with human scores for web UI similarity metrics. CC: Correlation coefficient (absolute value); OR: Outlier ratio; MAE: Mean absolute error; RMS: Root mean square error; SROCC: Spearman's rank-order correlation coefficient (absolute value). We mark the \textbf{best results} with bold font and the \underline{second best} with underline. Metrics are sorted by their SROCC.}
\label{tab:IQA}
\vspace{-5pt}
\begin{tabular}{@{}lcccccccccc@{}}
\toprule
 & \multicolumn{4}{c}{Variance-Weighted Regression} & \multicolumn{4}{c}{Non-Linear Regression} & \multicolumn{1}{c}{Direct} \\ 
\cmidrule(lr){2-5} \cmidrule(lr){6-9} \cmidrule(lr){10-10}
 & CC $\uparrow$ & MAE $\downarrow$ & RMS $\downarrow$ & OR $\downarrow$ & CC $\uparrow$ & MAE $\downarrow$ & RMS $\downarrow$ & OR $\downarrow$ & SROCC $\uparrow$ \\ 
\midrule
MAE   & \textbf{0.547} & \underline{4.10} & \textbf{1.95} & 0.049 & \underline{0.515} & \underline{0.646} & \underline{0.765} & 0.013 & \textbf{0.542} \\
NEMD  & \underline{0.469} & \textbf{3.98} & \textbf{1.95} & 0.052 & \textbf{0.532} & \textbf{0.628} & \textbf{0.752} & 0.023 & \underline{0.508} \\
PSNR  & 0.323 & 5.69 & 2.46 & \textbf{0.000} & 0.434 & 0.679 & 0.800 & 0.016 & 0.451 \\

CLIP  & 0.314 & 5.37 & \underline{2.41} & 0.013 & 0.426 & 0.681 & 0.800 & \underline{0.010} & 0.340 \\

SSIM  & 0.305 & 5.47 & 2.42 & \underline{0.010} & 0.381 & 0.699 & 0.817 & \textbf{0.000} & 0.218 \\

LPIPS & 0.221 & 5.70 & 2.49 & \underline{0.010} & 0.290 & 0.726 & 0.847 & 0.013 & 0.168 \\

\midrule
Human  &       &      &      &       &       &       &       &       & 0.640 \\
\bottomrule
\end{tabular}
\vspace{-5pt}
\end{table*}

\begin{table*}[h!]
    \centering
    \footnotesize
    \caption{Visual metrics of models across methods. SC: Self-contained; ZS: Zero-shot; CoT: Chain-of-thought; SR: Self-refine. The best result per model is highlighted in bold.}
    \label{tab:visual-metrics}
    \vspace{-5pt}
    \begin{tabular}{@{}lcccccccccccc@{}}
        \toprule
        Model & \multicolumn{4}{c}{Gemini-Pro} & \multicolumn{4}{c}{GPT-4o} & \multicolumn{4}{c}{Claude-3.5} \\
        \cmidrule(lr){2-5} \cmidrule(lr){6-9} \cmidrule(lr){10-13}
        Method & SC  & ZS  & CoT & SR & SC  & ZS  & CoT & SR & SC  & ZS  & CoT & SR \\
        \midrule
        MAE\textsubscript{\(\downarrow\)} & 68.3 & 66.7 & 66.5 & \textbf{63.1} & 69.3 & 67.7 & 67.9 & \textbf{65.1} & 71.6 & 69.1 & 69.3 & \textbf{69.4} \\
        NEMD\textsubscript{\(\uparrow\)} & 0.732 & 0.744 & 0.743 & \textbf{0.757} & 0.730 & 0.734 & 0.735 & \textbf{0.750} & 0.717 & \textbf{0.732} & 0.729 & 0.725 \\
        CLIP\textsubscript{\(\uparrow\)} & 0.771 & 0.785 & 0.789 & \textbf{0.797} & 0.753 & 0.766 & 0.766 & \textbf{0.774} & 0.767 & 0.788 & 0.784 & \textbf{0.792} \\
        \bottomrule
    \end{tabular}%
\vspace{-10pt}
\end{table*}

\begin{table}[h!]
    \centering
    \footnotesize
    \caption{RER of models across methods. SC: Self-contained; ZS: Zero-shot; CoT: Chain-of-thought; SR: Self-refine. The best result per model is highlighted in bold.}
    \label{tab:function-metrics}
    \vspace{-5pt}
    \begin{tabular}{@{}lcccc@{}}
        \toprule
         Act. Exist. Ratio$_\uparrow$ & SC    & ZS     & CoT    & SR \\
        \midrule
        Gemini-Pro   & 0.008 & 0.822 & 0.819 & \textbf{0.832} \\
        GPT-4o       & 0.006 & 0.779 & 0.774 & \textbf{0.803} \\
        Claude-3.5   & 0.007 & 0.640 & 0.615 & \textbf{0.668} \\
        \bottomrule
    \end{tabular}%
    \vspace{-10pt}
\end{table}

\begin{table*}[ht]
    \centering
    \footnotesize
    \caption{Comparison of metrics across different models and prompting strategies. ZS: Zero-shot; CoT: Chain-of-thought; SR: Self-refine. The best result per dimension is highlighted in bold.}
    \vspace{-5pt}
    \label{tab:finegrain}
    \begin{tabular}{@{}lccccccccc|c@{}}
        \toprule
        & \multicolumn{3}{c}{Gemini-Pro} & \multicolumn{3}{c}{GPT-4o} & \multicolumn{3}{c}{Claude-3.5} & \multirow{2}{*}{\textbf{Avg.}} \\
        \cmidrule(lr){2-4} \cmidrule(lr){5-7} \cmidrule(lr){8-10}
        \textbf{Difference}$_\downarrow$ & \textbf{ZS} & \textbf{CoT} & \textbf{SR} & \textbf{ZS} & \textbf{CoT} & \textbf{SR} & \textbf{ZS} & \textbf{CoT} & \textbf{SR} & \\
        \midrule
        \textbf{Pos. Shift}   & 0.239 & 0.260 & 0.236 & 0.217 & 0.267 & 0.209 & \textbf{0.155} & 0.161 & 0.193 & 0.215 \\
        \textbf{Area Diff}    & 0.376 & 0.389 & 0.403 & 0.360 & 0.375 & 0.360 & \textbf{0.279} & 0.326 & 0.334 & 0.356 \\
        \textbf{Color Diff}   & 0.128 & 0.144 & 0.110 & 0.043 & 0.066 & 0.055 & 0.057 & \textbf{0.024} & 0.038 & 0.074 \\
        \textbf{Text Diff}    & 0.014 & 0.011 & 0.009 & 0.013 & \textbf{0.004} & 0.005 & 0.020 & 0.011 & 0.005 & 0.010 \\
        \bottomrule
    \end{tabular}%
    \vspace{-10pt}
\end{table*}

\section{Experiment Results}
\subsection{The Best Web UI Similarity Metric}
\label{sec:rq1}
A critical challenge in the \taskname task is accurately evaluating web UI similarity. To verify the effectiveness of the evaluation metrics and determine the most suitable one, we initiated a human evaluation in the web UI domain, where we compared various image similarity methods (Section \ref{sec:metrics}) and discussed their alignment with human preferences. We sample 600 pairs of original and generated screenshots and recruit 14 college students with varying levels of familiarity with web applications to rate their perceived similarity on a Likert scale~\cite{joshi2015likert} within five categories: ``Highly Dissimilar'', ``Dissimilar'', ``Moderately Similar'', ``Similar'', and ``Highly Similar''. This setup follows standard image quality assessment (IQA) procedure~\cite{Wang2004ImageQA, VQEG2000}.

We analyzed the alignment between human judgments and objective similarity scores following established evaluation protocols~\cite{VQEG2000, Wang2004ImageQA}:

\begin{itemize}[leftmargin=*]
    \item Spearman Rank-Order Correlation Coefficient (SROCC): Measure the rank correlation between human and objective scores by capturing the consistency of the rankings.
    \item Variance-weighted Regression Analysis: Evaluate how well objective metrics predict human subjective scores using four key metrics: \textit{absolute correlation coefficient (CC)}, \textit{outlier ratio (OR)}, \textit{weighted mean absolute error (MAE)}, and \textit{weighted root mean square error (RMSE)}. This analysis identifies how well objective scores align with human judgments.
    \item Nonlinear Regression Analysis: Use logistic functions to model the relationship between objective and subjective scores when the relationship is monotonic but nonlinear. We compute \textit{CC}, \textit{OR}, \textit{MAE}, and \textit{RMSE} to measure the strength and quality of the alignment.
    \item Human Evaluators' Consistency: we calculated the SROCC for each pair of evaluators and averaged the results. This inter-rater reliability serves as a benchmark to evaluate the performance of objective metrics relative to human consensus.
\end{itemize}

Please refer to Appendix~\ref{appendix:IQA} for a detailed illustration of the metrics and evaluation practices.

\textbf{Pixel-based methods generally perform better. MAE and NEMD perform the best across all approaches} (Table~\ref{tab:IQA}). The result is initially counter-intuitive as most previous works adopt semantic-level and structure-level metrics such as CLIP and SSIM~\cite{Si2024Design2CodeHF, xiao2024interaction2code, zhou2024bridging}. However, this result aligns with the practice in frontend development, where developers use various tools~\cite{pixelparallel, perfectpixel} to overlay the design image and the result web page to compare their pixel-level differences to refine their UI code~\cite{pixelperfectmedium, pixelperfectsteps}. Variance-weighted regression results showed that MAE and NEMD achieved the highest correlation coefficients (CC: 0.547, 0.469) and lowest RMS (1.95), demonstrating strong predictive accuracy. In non-linear regression, NEMD excelled with the highest CC (0.532) and lowest MAE (0.628) and RMS (0.752), effectively capturing non-linear relationships. 

\textbf{Inter-rater reliability (SROCC) among human evaluators is 0.640, indicating a relatively high agreement and consistency in subjective assessments}. Among the metrics, MAE (0.542) and NEMD (0.508) demonstrated the best consistency with human rankings. 

\textbf{Despite their lower overall alignment scores, SSIM, CLIP, and LPIPS excelled in minimizing outliers (low OR)}, demonstrating their ability to reduce significant mismatches.

To further understand the characteristics of these metrics, we divide the image pairs into three equal-sized groups according to their human ratings (i.e., low, medium, high) and analyze their correspondence with human scores (details in Appendix~\ref{appendix:sim-analysis}). Analysis reveals pixel-based metrics, particularly MAE and NEMD, excel in low and medium-similarity cases. MAE is the most robust metric, maintaining strong performance across all similarity levels. While semantic and structural metrics, such as SSIM and LPIPS, perform better in high-similarity cases, they have near-zero performance in low and medium-similarity groups. This indicates that \textbf{pixel-level features are more effective at distinguishing dissimilar images, whereas semantic and structural information better captures fine-grained similarities.}

\subsection{Effectiveness of the Resource List}
\label{sec:rq2}
Central to our framework is a novel data structure, the resource list. To assess its impact, we employ MLLMs to generate web page code under various prompting strategies (Section~\ref{subsubsec:prompt}), using the two best-performing metrics (MAE and NEMD) and the best-performing high-level metrics (CLIP) for visual similarity and RER to measure the function similarity. We use the result of the self-contained (SC) prompt as a baseline. 

\textbf{Adding resource lists can improve the visual similarity of a generated webpage across different MLLMs and metrics}. Table~\ref{tab:visual-metrics} shows the comparison of visual metrics. We observe that SC consistently results in the lowest visual scores. This is because resource lists enable MLLMs to include the exact images displayed on the webpage, thus enhancing the overall similarity. Without resource lists, MLLMs can only use placeholder images in the generated web code. Some examples of such cases are in Appendix~\ref{appendix:visual-compare}, Fig.~\ref{fig:data-compare}. This highlights the practical value of resource lists in real-world web development compared to self-contained methods.

\textbf{Resource lists enable MLLMs to generate webpages with valid resources, significantly boosting RER from 0\% to 66\%-80\%}. Table~\ref{tab:function-metrics} shows that under SC prompting, MLLMs exhibit near-zero functional similarity due to the lack of guidance for generating valid links. However, some links are inferred through common sense (e.g., ``facebook.com'' for Facebook). Among prompting strategies, self-refine (SR) consistently achieves the highest scores across models, making it the most effective for both visual and functional metrics. Genimi-Pro emerges as the top-performing MLLM in reproducing functionality.

An interesting observation is that CoT prompting slightly decreases performance despite its reasoning capability. We discuss this phenomenon in Appendix~\ref{appendix:CoT}.

\subsection{Limitations of MLLMs in MRWeb}
\label{sec:rq3}
For all matched actionable element pairs, we calculate their fine-grained metrics to have a deeper understanding of MLLMs' strengths and weaknesses.

\textbf{The main challenge in \taskname generation is the visual grounding problem, where MLLMs struggle to replicate the position and size of elements} (Table~\ref{tab:finegrain}). This is reflected by the Positional Shift and Area Difference metrics. MLLMs generate elements with an average 21.5\% positional shift relative to the entire webpage and an average 35.6\% size difference compared to the original elements.

\textbf{Despite these issues, MLLMs perform well in recognizing color and text}, with Color Difference and Text Similarity metrics showing much smaller errors. Among the models, Claude-3.5 demonstrates the best positional accuracy and size recognition, with the lowest Positional Shift (15.5\%) and Area Difference (27.9\%) in the Zero-shot strategy. GPT-4o excels in Text Similarity (0.004), showing strong semantic accuracy. 

\begin{figure}[ht]
    \centering
        \begin{subfigure}[b]{0.49\linewidth}
            \includegraphics[width=\linewidth]{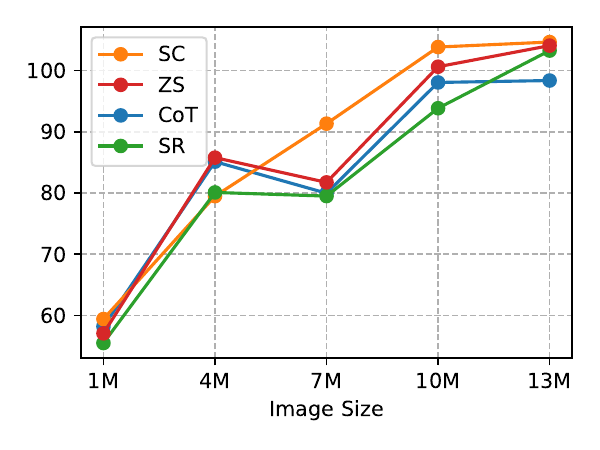}
            \vspace{-20pt}
            \caption{GPT-4o MAE$_\downarrow$}
            \label{fig:gpt4o-mae}
        \end{subfigure}
         \begin{subfigure}[b]{0.49\linewidth}
            \includegraphics[width=\linewidth]{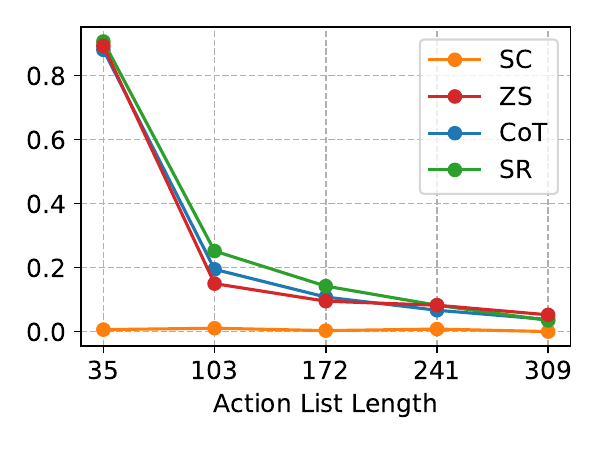}
            \vspace{-20pt}
            \caption{GPT-4o RER$_\uparrow$}
            \label{fig:gpt4o-aer}
        \end{subfigure}
        \vspace{-20pt}
        \caption{GPT-4o performance decreases with input image size (million pixels) and resource list length.}
        \vspace{-10pt}
\end{figure}

\textbf{The performance of MLLMs degrades with increasing input complexity, suggesting further optimization in handling large-scale and complex inputs.} To evaluate the robustness of MLLMs in the \taskname task, we analyzed their visual (MAE) and functional (RER) performances across varying image sizes and resource list lengths. As shown in Fig. \ref{fig:gpt4o-mae} and \ref{fig:gpt4o-aer}, GPT-4o's performance lowers with increasing input complexity. While it performs well on simpler websites with smaller image sizes and shorter resource lists, its accuracy declines significantly for intricate websites. Similar patterns are observed in other MLLMs, as detailed in Appendix~\ref{appendix:complexity}.

\subsection{\taskname’s Practical Capabilities.}
\label{sec:rq4}
We developed a user-friendly tool within the \taskname framework to convert visual designs into multi-page, realistic web UI code. The tool's interface is shown in Appendix~\ref{appendix:demo}. We conducted a case study using AI-generated design images\footnote{\url{https://openai.com/index/dall-e-3/}} to build a personal website with three pages: home, project, and contact (Appendix~\ref{appendix:demo}). We introduced various resources to each page to test the tool’s capabilities. These resources include internal links, external links, images, and backend routing (Table~\ref{tab:practical}). The tool successfully addressed all the challenges, achieving a 100\% success rate. The demonstration video of the entire development procedure is available online\footnote{\url{https://github.com/WebPAI/MRWeb}}. Specifically, the generated home page and project page included internal and external links and embedded images with pixel-perfect alignment. The contact page demonstrated the tool’s ability to integrate backend routing seamlessly, implying its full-stack capabilities. 

\begin{table}[ht]
    \centering
    \footnotesize
    \caption{Total number of challenges in the home, project, and contact page versus tool success rate.}
    \vspace{-5pt}
    \label{tab:practical}
    \begin{tabular}{lcc}
        \toprule
        Resource & Count & Success \\
        \midrule
        Internal links &  7 & 100\% \\
        External links &  9 & 100\% \\
        Images         &  13 & 100\% \\
        Backend routing & 1 & 100\% \\
        \bottomrule
    \end{tabular}
    \vspace{-5pt}
\end{table}

\section{Discussions}

This section highlights key implications of our work for future research.

\paragraph{Visual metrics for UI quality (RQ1)} While prior studies emphasize structural and semantic metrics, our findings show that pixel-based metrics better align with human judgment, especially in low-to-medium similarity cases. This suggests hybrid approaches that combine these metrics could provide more robust evaluations. The limited performance of learning-based methods in these cases indicates a need for targeted fine-tuning.

\paragraph{Enhancing website generation with resource lists (RQ2)} Incorporating resource lists significantly boosts both visual and functional metrics, underscoring their potential for advancing automated full-stack development.

\paragraph{Improving MLLM visual grounding (RQ3)} Metrics on positional shift and area difference highlight MLLMs’ limitations in precise positioning and sizing. Addressing this may involve improving visual grounding or developing layout-aware prompts for better layout reproduction.

\paragraph{Advancing MRWeb generation (RQ4)} MRWeb generation connects design and functionality, supporting links, images, and routing. However, non-link-based functionalities remain underexplored, presenting opportunities for more comprehensive full-stack development.

\section{Conclusion}
In this paper, we introduce the MRWeb generation task, addressing the limitations of single-page design-to-code methods. Our contributions include defining the MRWeb problem, creating a benchmark dataset, conducting a comprehensive IQA for web UIs, analyzing MLLM performance, and developing a dedicated MRWeb generation tool. We release the tool, dataset, and evaluation framework to facilitate future research.
\newpage

\section*{Limitations}

\textit{Limited Support for Non-Link Functionalities.}
While MRWeb effectively handles links, images, and routing, it does not currently support non-link-based functionalities, as these require distinct formulations and evaluation metrics. Addressing this limitation is a key focus for future work, with the goal of enabling full-stack development capabilities.

\noindent\textit{Context Length Constraints.}
MLLMs have limited context windows (e.g., 128K tokens for GPT-4o), which can be a challenge for websites with extensive token requirements. However, our experiments show that all prompts remain within these limits, highlighting the approach's feasibility for most practical scenarios.

\noindent\textit{Backbone Model Selection.}
We validate \taskname using three popular multimodal LLMs, but smaller models struggle with complex prompts. To improve adaptability and generalization, future work will explore the potential of emerging models and investigate strategies to handle more complex input scenarios.

\bibliographystyle{acl_natbib}
\bibliography{reference}

\appendix

\section{Motivating study}
\label{sec:motivating-study}

\begin{figure*}[t]
    \begin{minipage}{0.3\textwidth}
        \footnotesize
        \centering
        \begin{tabular}{lccc}
        \toprule
        \textbf{Stat} & \textbf{Ext./Back.} & \textbf{Int.} & \textbf{Img} \\
        \midrule
        \textbf{Median}   & 26  & 22   & 14 \\
        \textbf{Mean}  & 51  & 45  & 25 \\
        \textbf{Max}   & 361 & 763 & 537 \\
        \bottomrule
        \end{tabular}
        \caption{Summary statistics for external/backend links, internal links, and images count.}
        \label{fig:motivating-table}
    \end{minipage}
    \begin{minipage}{0.7\textwidth}
        \centering
        \begin{subfigure}[b]{0.4\textwidth}
            \centering
            \includegraphics[width=\linewidth]{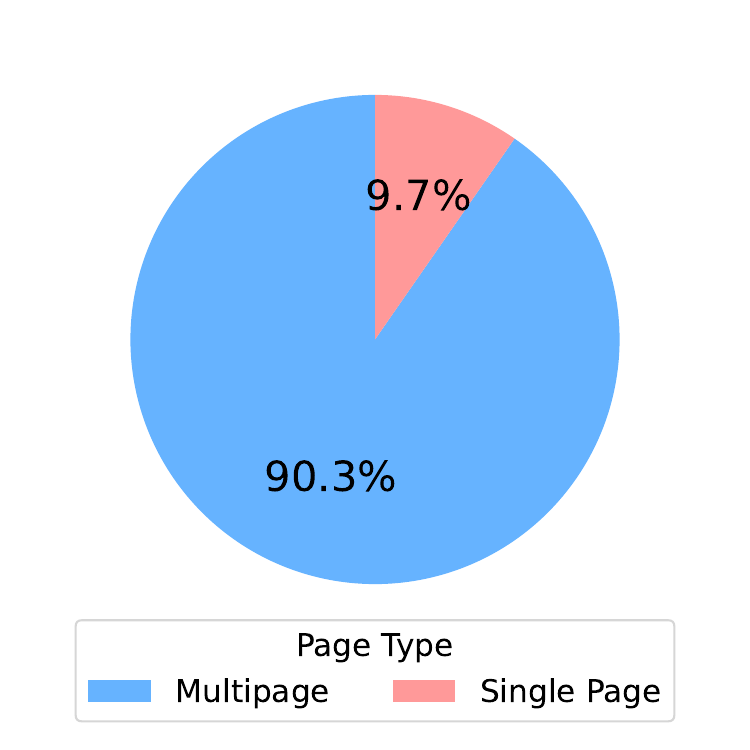}
            \caption{Proportion of multi-page vs single-page websites.}
            \label{fig:multipage-pie}
        \end{subfigure}
        \begin{subfigure}[b]{0.4\textwidth}
            \includegraphics[width=\linewidth]{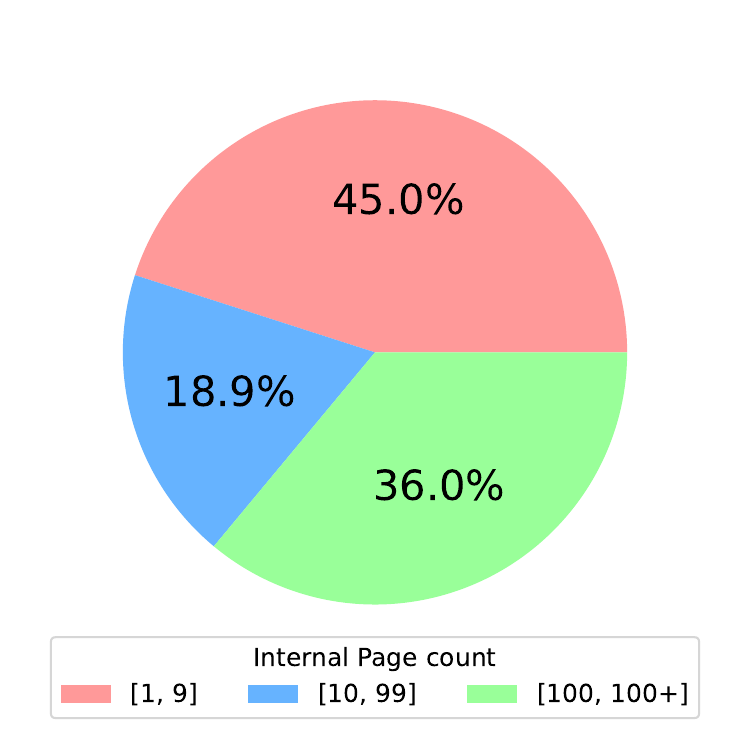}
            \caption{Number of internal pages in multi-page websites.}
            \label{fig:internal-page-pie}
        \end{subfigure}

        \caption{Detailed analysis.}
        \label{fig:RQ1-detail}
    \end{minipage}
\end{figure*}

In this study, we show that the three common assumptions made by previous design-to-code works (i.e., UIs are self-contained and do not have external links, UIs consist of a single or a limited number of pages, and UIs use placeholder images) are highly over-simplified in web UI scenarios. We sample the top 300 most visited websites ranked by the Tranco~\footnote{https://tranco-list.eu} list and study their characteristics. 

Specifically, we first count the number of external links, internal links, and images on each web page. The statistics in Figure \ref{fig:motivating-table} highlight the over-simplified assumptions about web UIs: The number of pages involved is typically large, with a mean of 51 external links, 45 internal links, and 25 images. The maximum number of external links reaches 361, internal links 763, and images 537, indicating that websites generally contain numerous elements that are not accounted for in simplified models. 

Then, we calculate the proportion of multi-page websites among the samples. Figure \ref{fig:multipage-pie} illustrates the result, which shows that 90.3\% of the websites are multi-page, while only 9.7\% are single-page websites. This suggests that most modern websites are not self-contained but involve a complex structure with multiple pages. 

Finally, we calculate the number of distinct internal pages each multi-page website contains. Internal pages are web pages within the same domain as the website; for instance, "apple.com/iphone" and "apple.com/ipad" would both be internal to the website "apple.com," while "facebook.com" is external to the website. To count the number of internal pages for each website, we recursively visit all the internal pages a website links to until no new internal web pages can be found. The result is shown in Figure \ref{fig:internal-page-pie}. Less than 50\% of the multi-page websites have fewer than 10 internal pages, while 18.9\% of the websites have between 10 and 99 internal pages, demonstrating a moderate complexity. Notably, 36\% of the websites feature more than 100 internal pages, showcasing the highly complex nature of a significant portion of modern websites. This variability in the number of internal pages challenges the assumption that web UIs consist of a single or a limited number of pages.

\begin{tcolorbox}[colback=gray!20, colframe=gray!20, width=\columnwidth]
\textbf{Conclusion:} The three common assumptions made by previous design-to-code works are over-simplified, and manually locating and editing the omitted content in generated code requires high proficiency in coding and remarkable effort.
\end{tcolorbox}

\section{Related Works}
\label{appendix:related-work}
Current approaches of design-to-code span deep learning (DL), computer vision (CV), and multimodal large language models (MLLMs). DL-based methods use CNNs for GUI prototyping \cite{acsirouglu2019automatic, Cizotto2023WebPF, Moran2018MachineLP, Xu2021Image2e, Chen2018FromUI}, with Pix2code \cite{beltramelli2018pix2code} combining CNNs and LSTMs for DSL generation, and \cite{Chen2022CodeGF} enhancing quality via attention-based encoder-decoder frameworks. In CV, Sketch2Code \cite{jain2019sketch2code} processes hand-drawn sketches using object detection, while REMAUI \cite{nguyen2015reverse} employs OCR to extract UI elements and build hierarchies. MLLM-based methods offer more advanced multi-modal generation. Design2Code \cite{Si2024Design2CodeHF} uses text-augmented prompts, DCGen \cite{wan2024automatically} employs a divide-and-conquer layout approach, DeclarUI \cite{zhou2024bridging} builds page transition graphs for multi-screen apps, and Interaction2Code \cite{xiao2024interaction2code} generates interactive UIs from web interaction graphs. Despite these advances, most methods still rely on the three over-simplified assumptions, leaving the challenge of multi-page website generation underexplored.

\section{Prompts}
\label{appendix:prompts}
In this section, we list the exact prompts used in the experiments.

\paragraph{Self-contained prompt.}\texttt{Here is a screenshot of a web page. Please write an HTML and Tailwind CSS to make it look exactly like the original web page. Pay attention to things like size, text, position, and color of all the elements, as well as the overall layout. Respond with the content of the HTML+tail-wind CSS code.}

\paragraph{Zero-shot prompt.} \texttt{Here is a screenshot of a web page and its ``action list'' which specifies the links and images in the webpage. Please write an HTML and Tailwind CSS to make it look exactly like the original web page. Pay attention to things like size, text, position, and color of all the elements, as well as the overall layout. The format of the action list is as follows:
    \{
    ``position'': bounding box of format [[x1, y1], [x2, y2]], specifying the top left corner and the bottom right corner of the element;
    ``type'': element type;
    ``url'': url of the element;
    \}
The action list is as follows: 
[ACTION LIST]}

\paragraph{CoT prompt.} \texttt{Here is a screenshot of a web page and its ``action list'' which specifies the links and images in the webpage. Please write a HTML and Tailwind CSS to make it look exactly like the original web page. Please think step by step, and pay attention to things like size, text, position, and color of all the elements, as well as the overall layout.  The format of the action list is as follows:
    \{
    ``position'': bounding box of format [[x1, y1], [x2, y2]], specifying the top left corner and the bottom right corner of the element;
    ``type'': element type;
    ``url'': url of the element;
    \}
The action list is as follows:
[ACTION LIST]}

\paragraph{Self-refine prompt.} \texttt{Here is a screenshot of a web page and its ``action list'' which specifies the links and images in the webpage. I have an HTML file for implementing a webpage but it has some missing or wrong elements that are different from the original webpage. Please compare the two webpages and revise the original HTML implementation. Return a single piece of HTML and tail-wind CSS code to reproduce exactly the website. Pay attention to things like size, text, position, and color of all the elements, as well as the overall layout. Respond with the content of the HTML+tail-wind CSS code.  The format of the action list is as follows:
    \{
    ``position'': bounding box of format [[x1, y1], [x2, y2]], specifying the top left corner and the bottom right corner of the element;
    ``type'': element type;
    ``url'': url of the element;
    \}
The current implementation I have is: [CODE] The action list is as follows: [ACTION LIST]}

\section{Visual Similarity Metrics}
\label{appendix:similarity}
This section provides all the details of the visual similarity metrics tested in this work.

At the pixel level, we employ metrics that directly compare pixel values to quantify low-level differences: 
\begin{itemize}[leftmargin=*] 
\item \textbf{Mean Absolute Error (MAE)}~\cite{Nguyen2015ReverseEM, Moran2018MachineLP}: Computes the average absolute difference in pixel intensities, providing a straightforward measure of overall similarity that treats all errors equally without amplifying larger differences.

\item \textbf{Peak Signal-to-Noise Ratio (PSNR)}~\cite{Lim2017EnhancedDR, Wang2019EDVRVR}:  Measures the ratio between the maximum possible power of a signal (image) and the power of corrupting noise. PSNR is based on the Mean Squared Error (MSE), with higher values indicating closer similarity between two images. It is widely used to evaluate image quality, especially in compression and restoration tasks. 

\item \textbf{Wasserstein Distance (Earth Mover's Distance - EMD)}~\cite{Arjovsky2017WassersteinG, Rubner2000TheEM}: Measures the minimum transport cost required to transform one image onto the other, capturing spatial differences in pixel values and reflects structural rearrangements needed to align images. The EMD depends on the image size, with larger image pairs producing higher EMD values due to more pixels. To eliminate this dependency and make the metric size-independent, we define a normalized version $\text{\textbf{NEMD}} = 1 - \frac{\text{EMD}}{\text{EMD}_{\text{max}}}$ where $\text{EMD}_{\text{max}}$ is the maximum possible EMD between a reference image and any other arbitrary image. It represents the worst-case scenario of pixel differences (i.e., the distance from the reference image to a "completely" different image). It is achieved by assuming each pixel in the reference image is moved to its farthest possible value (0 or 255), thus providing an upper bound for the EMD. The NEMD ranges from 0 to 1, where higher values indicate greater similarity. This metric is not symmetric, as the $\text{EMD}_{\text{max}}$ is computed relative to the reference image. The original web page screenshot serves as the reference image in our experiments.

\end{itemize}

At the structural level, we use metrics that capture spatial and perceptual coherence: 
\begin{itemize}[leftmargin=*] 

\item \textbf{Structural Similarity Index Measure (SSIM)}~\cite{Zhou2024BridgingDA, Wang2004ImageQA}: Assesses structural coherence by evaluating changes in luminance, contrast, and structural information across images. SSIM models perceived image quality by accounting for local patterns and how structural details align, closely mirroring human visual perception. 

\end{itemize}

At the semantic level, we leverage: \begin{itemize}[leftmargin=*] 

\item \textbf{CLIP score}~\cite{Radford2021LearningTV, Si2024Design2CodeHF}: Derived from the CLIP model, it captures high-level semantic similarities by aligning image embeddings with corresponding language representations. This approach allows us to gauge similarity based on shared meanings and conceptual elements rather than visual appearance alone. It is particularly suited for comparing images representing similar objects or scenes in different styles or contexts. 

\item \textbf{Learned Perceptual Image Patch Similarity (LPIPS)}~\cite{Zhang2018TheUE}: Evaluates perceptual similarity by computing the distance between deep feature representations of two images. LPIPS offers a robust method for assessing how similar two images are regarding human perception. We use VGG~\cite{Simonyan2014VeryDC} as the backbone model for LPIPS calculation.

\end{itemize}

Except for the CLIP score, which accepts images of varying sizes as input, we pad each image pair with random noise to ensure the two images in the pair are the same size.

\section{Image Quality Analysis Details}
\label{appendix:IQA}
We analyzed the alignment between human judgments and objective similarity scores following established evaluation protocols~\cite{VQEG2000, Wang2004ImageQA}. This section provides a detailed illustration of the metrics and evaluation practices.

\paragraph{Processing Human Scores} 
After collecting human-perceived similarity scores, the scores of each annotator across all image pairs were normalized using z-scores (mean-centered and scaled by standard deviation) to ensure comparability. We computed the Mean Opinion Scores (MOS) for each image pair by first removing outliers and then averaging all human scores of the image pair. This process yields a robust, representative set of subjective human scores. The image database and subjective scores will be publicly available for further study.

\begin{table*}[]
\centering
\small
\caption{Correlation of similarity scores with human scores across different score levels (Low, Medium, High). CC-V: Correlation coefficient under variance-weighted regression; CC-N: Correlation coefficient under non-linear regression; SROCC: Spearman's rank-order correlation coefficient. All correlations are absolute values. We mark the \textbf{best results} with bold font and the \underline{second best} with underline. \textbf{SROCC is a more direct measurement} of correlations between similarity scores and human scores.}
\label{tab:level-IQA}
\begin{tabular}{@{}lccc|ccc|ccc@{}}
\toprule
 & \multicolumn{3}{c|}{Low} & \multicolumn{3}{c|}{Medium} & \multicolumn{3}{c}{High} \\
\cmidrule(lr){2-4} \cmidrule(lr){5-7} \cmidrule(lr){8-10}
 & CC-V & CC-N & SROCC & CC-V & CC-N & SROCC & CC-V & CC-N & SROCC \\
\midrule
MAE & \underline{0.413} & \underline{0.429} & \underline{0.401} & \textbf{0.211} & \underline{0.222} & \textbf{0.212} & 0.406 & 0.420 & 0.330 \\
NEMD & \textbf{0.422} & \textbf{0.430} & \textbf{0.408} & 0.075 & 0.108 & \underline{0.209} & 0.286 & 0.301 & 0.328 \\
PSNR & 0.264 & 0.313 & 0.351 & \underline{0.193} & 0.192 & 0.191 & \textbf{0.582} & \textbf{0.583} & 0.264 \\
CLIP & 0.065 & 0.194 & 0.058 & 0.143 & \textbf{0.262} & 0.145 & 0.357 & 0.376 & 0.361 \\
SSIM & 0.033 & 0.033 & 0.012 & 0.066 & 0.069 & 0.058 & \underline{0.494} & \underline{0.508} & \textbf{0.379} \\
LPIPS & 0.041 & 0.044 & 0.084 & 0.046 & 0.049 & 0.010 & 0.491 & 0.502 & \underline{0.367} \\
\bottomrule
\end{tabular}
\end{table*}

\paragraph{Alignment Analysis of Objective and Subjective Scores} To compare the correlation between automatic objective scores and human subjective scores, we use three approaches:

\begin{itemize}[leftmargin=*]
\item \textbf{Spearman rank-order correlation coefficient (SROCC):} A direct measure of the strength and direction of the association between the ranked objective and subjective scores. In our experiment, we consider the absolute value of SROCC, which is a metric between 0 and 1, where a indicates perfect positive or negative correlation, and 0 indicates no correlation. SROCC is particularly useful for evaluating the consistency of scores produced by objective metrics with those of human subjective assessment scores, regardless of the absolute values of the scores.

\item \textbf{Variance-weighted regression analysis:} This analysis evaluates how well the objective metrics can predict human subjective scores. By weighting data points based on their variance before conducting a linear regression, we minimize the influence of noisy or uncertain scores, which leads to more robust and reliable predictions. After regression, we calculate several quantities to measure the magnitude of prediction errors and how closely the objective metric matches human assessments. The metrics are 1) the \textit{absolute correlation coefficient (CC)}: after taking the absolute value, this becomes a score between [0, 1] that measures the linear alignment between predictions and human perceptions, with higher absolute values indicating stronger alignment; 2) the \textit{outlier ratio (OR)}, i.e., the percentage of predictions outside twice the standard deviation, which identifies the ratio of cases where the model significantly deviates from human perception; 3) weighted \textit{mean absolute error (MAE)}, which measures the average magnitude of prediction errors without considering their direction, providing a straightforward indication of overall prediction accuracy; and 4) weighted \textit{root mean square error (RMSE)}, which gives a higher penalty to larger errors and emphasizes the magnitude of extreme deviations. These metrics collectively provide a comprehensive evaluation of the model’s ability to predict subjective human scores accurately and robustly.

\item \textbf{Nonlinear regression analysis:} In this analysis, logistic functions are used to fit a nonlinear mapping between objective and subjective scores. This approach is particularly useful when the relationship between objective metrics and human perception is monotonic but not strictly linear. We also calculate \textit{CC}, \textit{OR}, \textit{MAE}, and \textit{RMSE} to understand the strength of the nonlinear relationship and discrepancies between the metrics and human evaluations.

\end{itemize}

\paragraph{Assessing inter-rater reliability of human scores.} For each pair of human subjects, we calculate their \textbf{SROCC} to evaluate the consistency of their subjective assessments. We then average the SROCC values across all pairs to obtain an overall measure of inter-rater reliability. This indicates how consistently human subjects rank the similarity between objective and subjective scores, serving as a benchmark for evaluating the performance of objective metrics.

\section{Analysis of Similarity Metrics Across Different Similarity Levels}
\label{appendix:sim-analysis}

\paragraph{Why do learning- and structure-based metrics fail?} We divide the image pairs into three equal-sized groups according to their human ratings (i.e., low, medium, high) and analyze their correspondence with human scores. The result is in Table~\ref{tab:level-IQA}. 

For the low-similarity image group, NEMD consistently achieves the best results (CC-V: 0.422, CC-N: 0.430, SROCC: 0.408), followed by MAE. Notably, SSIM, CLIP, and LPIPS show near-zero correlation with human perception in this range.

In the medium-similarity image group, the performance of all metrics generally declines, indicating increased difficulty in predicting similarity when human perception becomes less polarized. NEMD and MAE remain competitive, and CLIP performs best under a nonlinear mapping with a CC-N  of 0.262.

For the high-similarity image group, PSNR achieves the highest correlation under variance-weighted mapping and non-linear mapping  (CC-V: 0.582, CC-N: 0.583). Metrics such as SSIM and LPIPS show significant improvement. SSIM achieves the highest direct correlation (SROCC: 0.379), while LPIPS also demonstrates notable performance in SROCC (0.367). 

In conclusion, \textbf{learning- and structure-based metrics fall short due to near-zero performance in low-similarity groups and poor performance in medium-similarity groups, failing to capture the similarity of intrinsically dissimilar image pairs.} In contrast, pixel-based metrics effectively distinguish dissimilar pairs while maintaining reasonable performance for high-similarity pairs, with MAE and NEMD emerging as standout methods.

\section{Why CoT Decrease Performance}
\label{appendix:CoT}
Manual investigation reveals that MLLMs tend to omit content at the end of their generated code, often ending with placeholder comments like ``additional content goes here.'' The decrease in CoT performance arises because when being explicitly prompted with ``please think step by step,'' it generates HTML code part by part, omitting content at the end of every part, thereby compounding the omissions compared to direct prompts. This leads to lower action existence rates and reduced overall performance.

\section{MLLMs' Performance and Input Complexity}
\label{appendix:complexity}
\begin{figure*}[ht]
    \centering
        \begin{subfigure}[b]{0.32\textwidth}
            \centering
            \includegraphics[width=\linewidth]{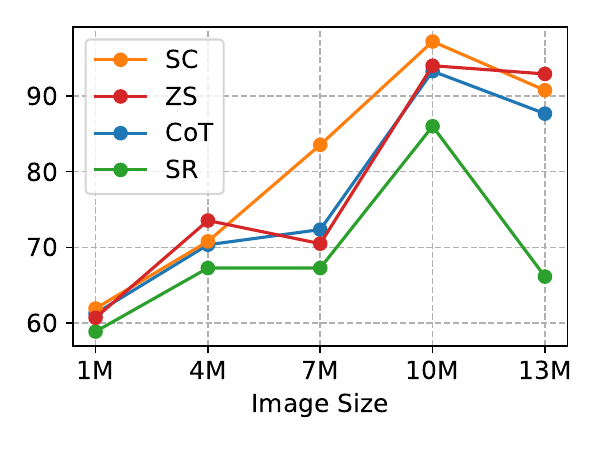}
            \caption{Gemini-Pro-1.5.}
        \end{subfigure}
        \begin{subfigure}[b]{0.32\textwidth}
            \includegraphics[width=\linewidth]{Sections/figs/gpt4o_MAE.pdf}
            \caption{GPT-4o.}
        \end{subfigure}
         \begin{subfigure}[b]{0.32\textwidth}
            \includegraphics[width=\linewidth]{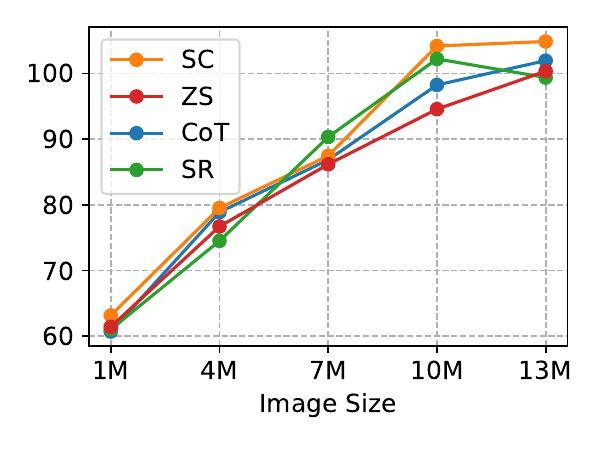}
            \caption{Claude-3.5.}
        \end{subfigure}
        \caption{MAE$_\downarrow$ vs. input image size (million pixels).}
        \label{fig:mae}
\end{figure*}

\begin{figure*}[ht]
    \centering
        \begin{subfigure}[b]{0.32\textwidth}
            \centering
            \includegraphics[width=\linewidth]{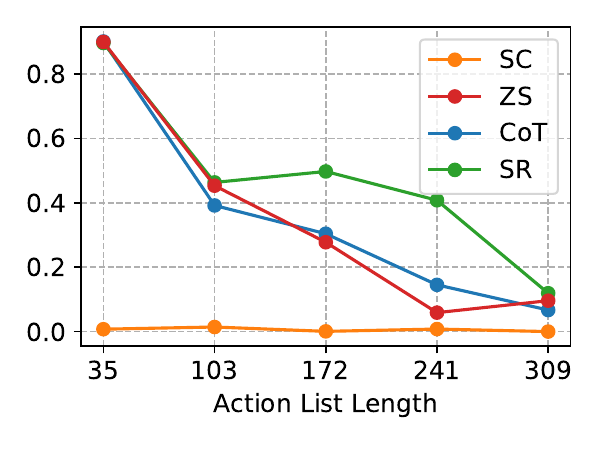}
            \caption{Gemini-Pro-1.5.}
        \end{subfigure}
        \begin{subfigure}[b]{0.32\textwidth}
            \includegraphics[width=\linewidth]{Sections/figs/gpt4o_match_ratio.pdf}
            \caption{GPT-4o.}
        \end{subfigure}
         \begin{subfigure}[b]{0.32\textwidth}
            \includegraphics[width=\linewidth]{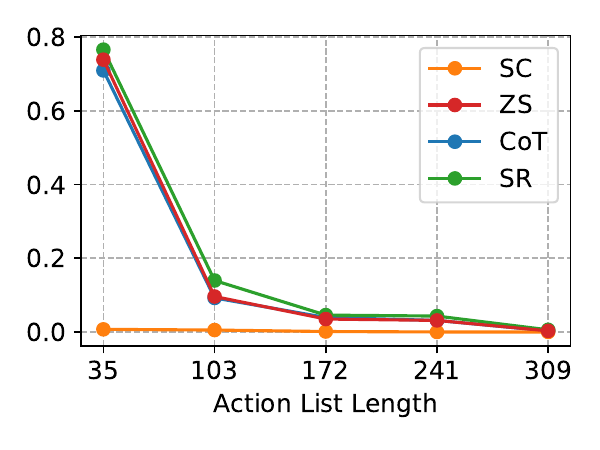}
            \caption{Claude-3.5.}
        \end{subfigure}
        \caption{Action existence ratio$_\uparrow$ vs. action list length.}
        \label{fig:aer}
\end{figure*}

According to Fig.~\ref{fig:mae}, as input image size increases, the MAE ($\downarrow$) grows for all models. GPT-4o and Claude-3.5 demonstrate more stable performance compared to Gemini-Pro, which is sensitive to size, particularly under the self-refine strategy. Among the prompting strategies, self-refine consistently performs best for Gemini-Pro and GPT-4o. For Claude-3.5, self-refine shows minimal improvement and occasionally underperforms compared to zero-shot and CoT).

For action list length (Fig.~\ref{fig:aer}), the action existence ratio ($\uparrow$) decreases as the list grows longer, highlighting the challenge of maintaining accuracy with increased complexity. GPT-4o and Claude-3.5 perform comparably, with CoT and self-refine providing slight advantages. However, for Claude-3.5, SR again offers a limited improvement to other methods.

\section{Developing Website with the MRWeb Tool}
\label{appendix:demo}
\begin{figure}[H]
    \centering
    \includegraphics[width=\linewidth]{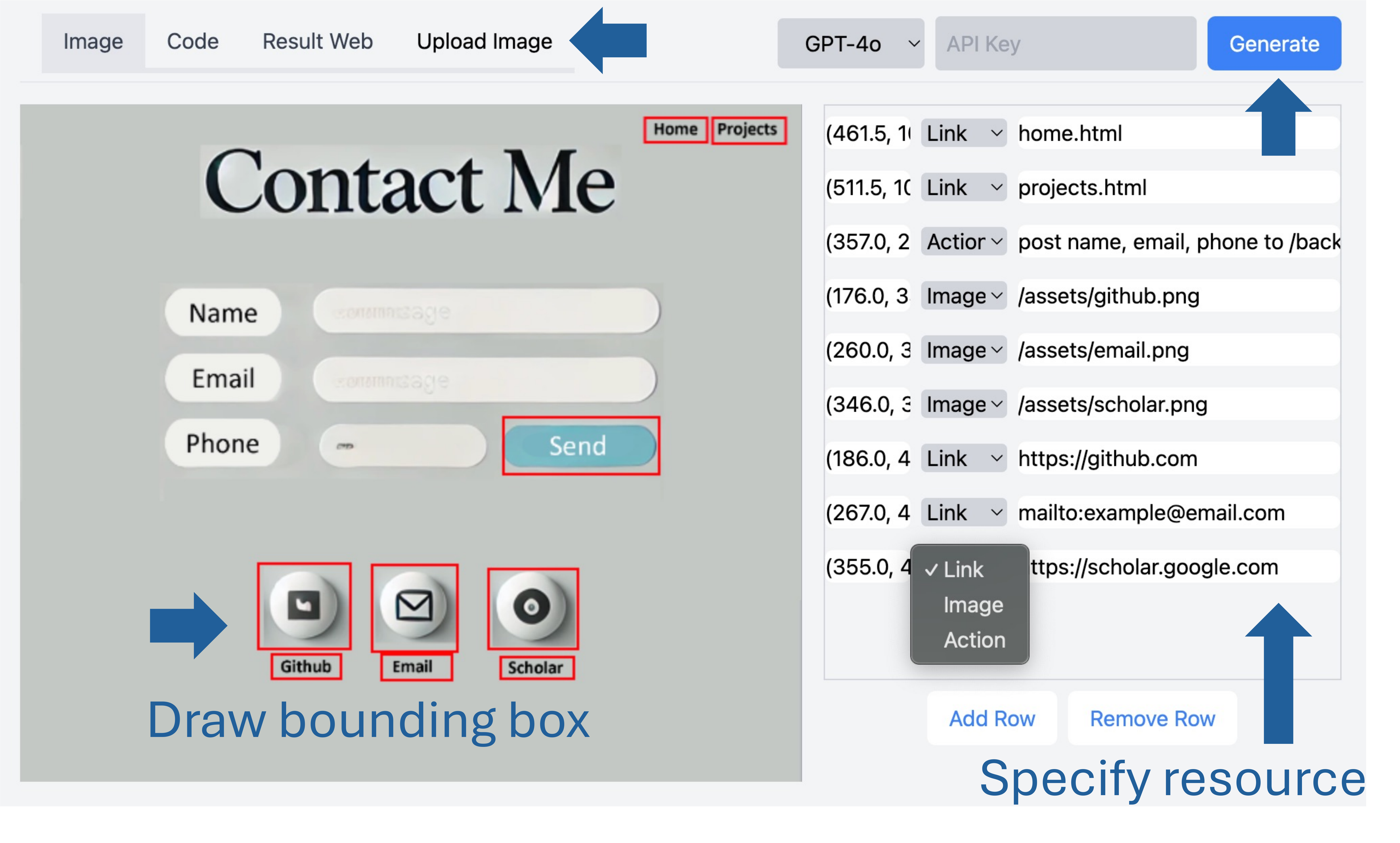}
    \caption{User interface of the MRWeb tool.}
    \label{fig:tool}
\end{figure}

In RQ4, we conduct a case study using the MRWeb tool, whose user interface is shown in Figure~\ref{fig:tool}. The tool enables users to define design assets, including webpage layouts, external resources (like images and links), and actions such as backend routing. These assets are organized using the action list structure—a dictionary-like format that systematically maps resources to visual elements. Leveraging the \taskname framework, the tool processes action lists and screenshots as inputs to MLLMs, facilitating the generation of functional, multi-page web UI code with visual consistency.

\begin{figure}[H]
    \centering
    \begin{subfigure}[b]{0.28\columnwidth}
            \centering
            \includegraphics[width=\linewidth]{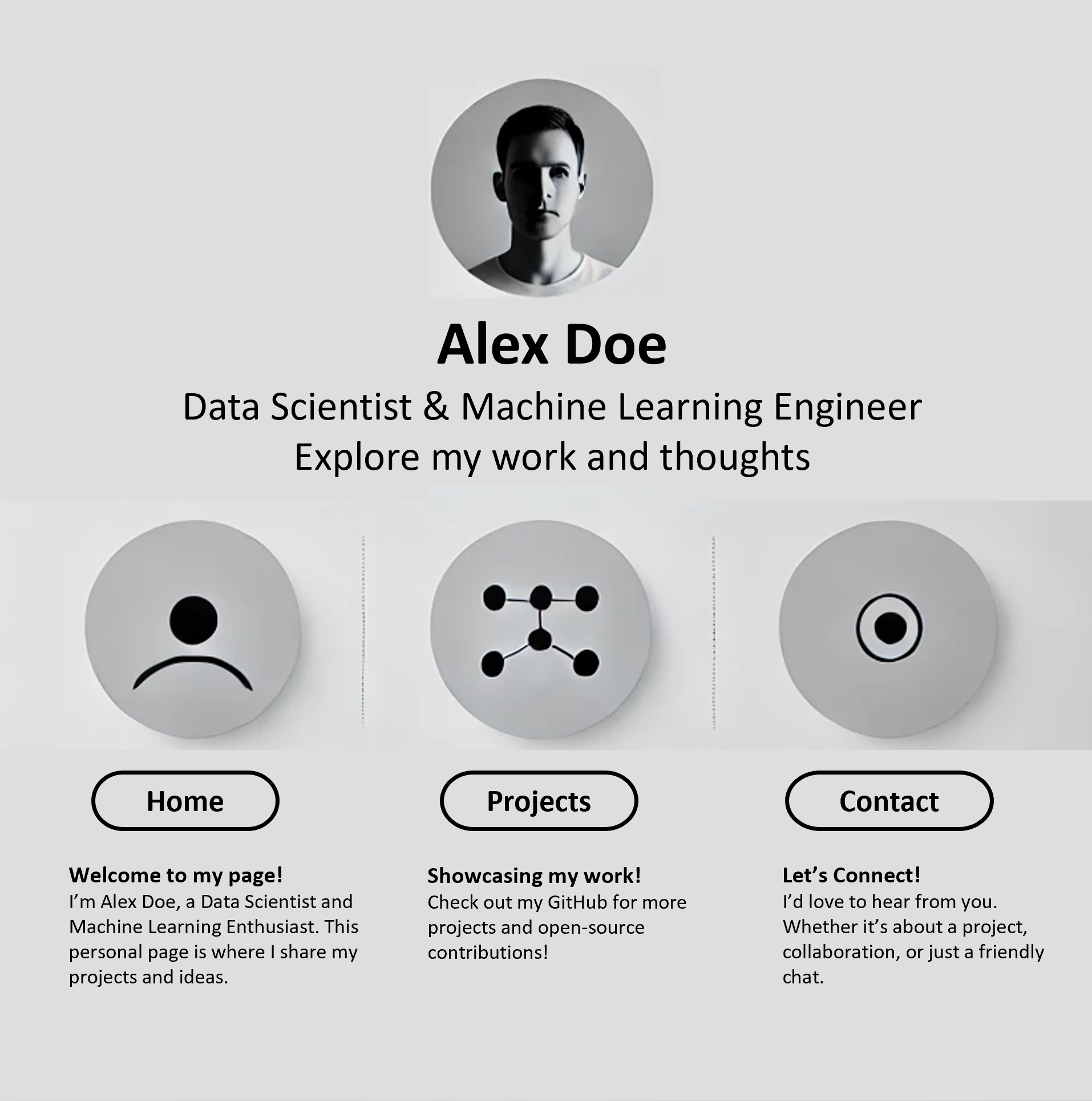}
            \caption{Home page design.}
        \end{subfigure}
        \begin{subfigure}[b]{0.294\columnwidth}
            \includegraphics[width=\linewidth]{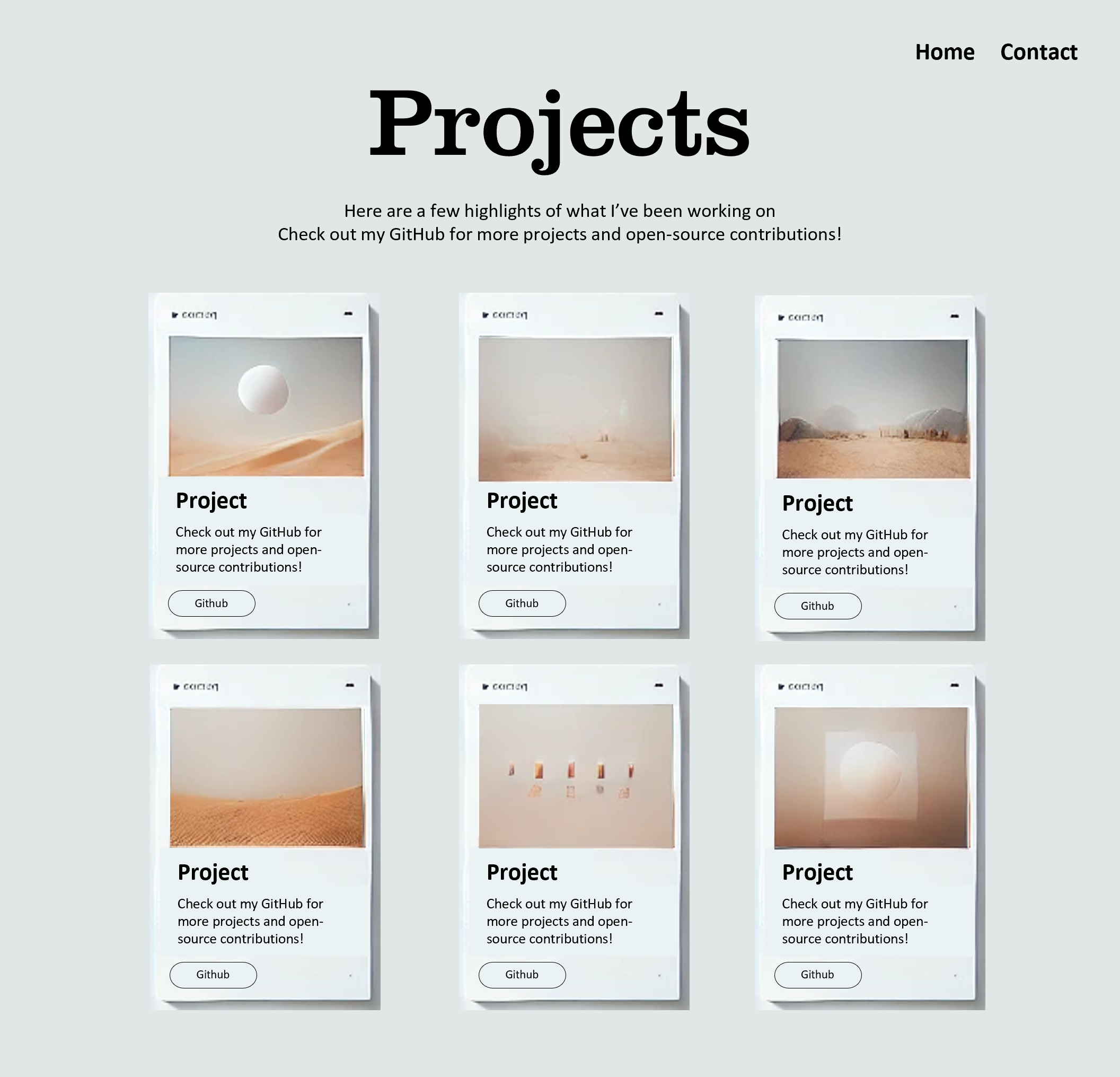}
            \caption{Project page design.}
        \end{subfigure}
         \begin{subfigure}[b]{0.312\columnwidth}
            \includegraphics[width=\linewidth]{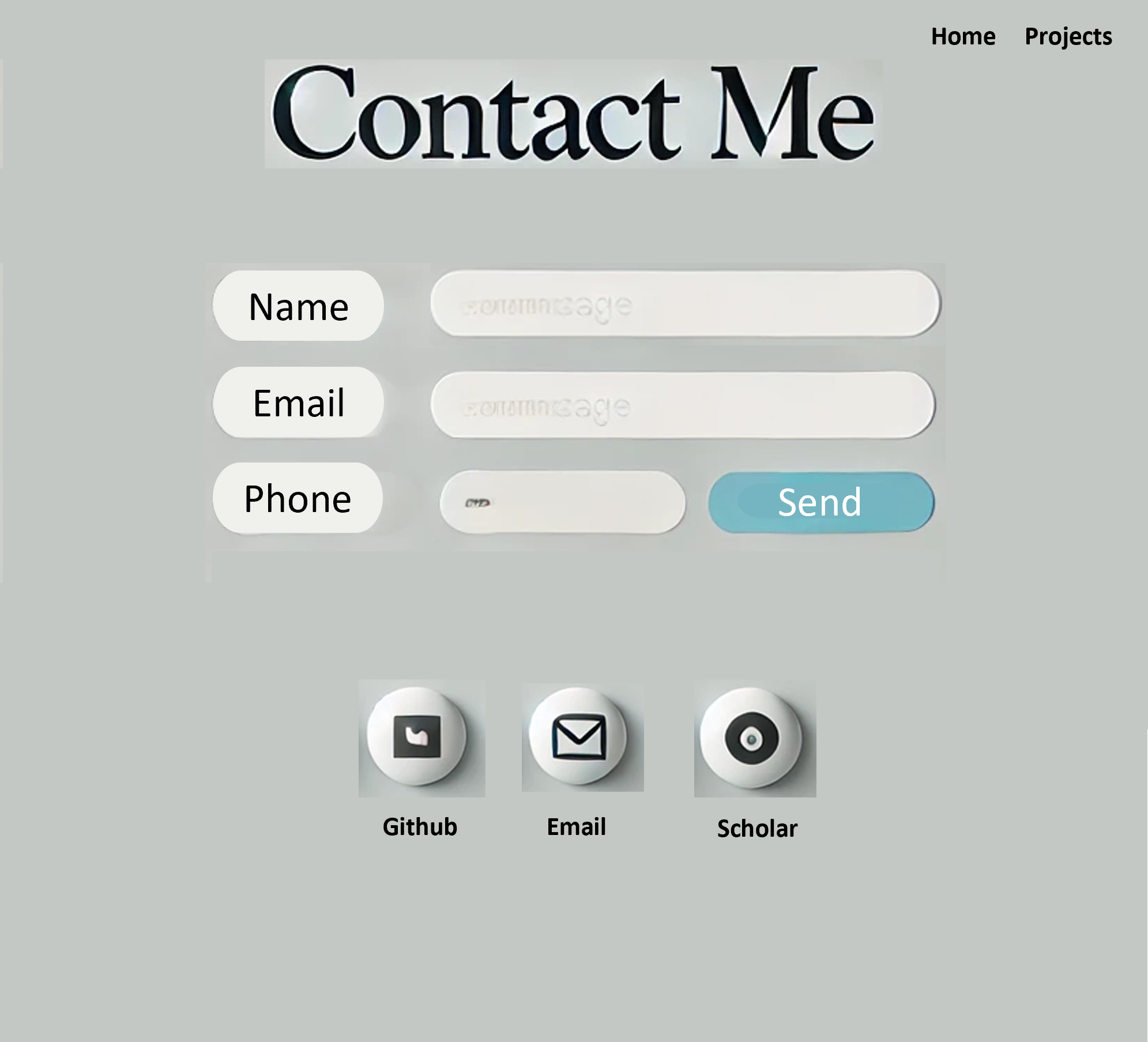}
            \caption{Contact page design.}
        \end{subfigure}
        \caption{Design images of the personal website in RQ4.}
        \label{fig:demo}
\end{figure}

The case study focuses on building a personal website with three internal pages: a home page, a project page, and a contact page, using AI-generated design images (Fig~\ref{fig:demo}). Table~\ref{tab:challenge} summarizes the key challenges introduced for each page.

\begin{table}[H]
    \centering
    \footnotesize
    \caption{Challenges and performance of the case study.}
    \label{tab:challenge}
    \begin{tabular}{llcc}
        \toprule
        & Challenge      & Count & Success \\
        \midrule
        Home page & Internal links & 3 & 100\% \\
                           & Images                 & 4 & 100\% \\
        Project page & Internal links       & 2 & 100\% \\
                           & External links & 6 & 100\% \\
                           & Images         & 6 & 100\% \\
        Contact page & Internal links       & 2 & 100\% \\
                           & External links         & 3 & 100\% \\
                           & Images                & 3 & 100\% \\
                           & Backend routing & 1 & 100\% \\
        \bottomrule
    \end{tabular}
\end{table}

\section{Visual Comparison of Self-Contained Web and MRWeb}
\label{appendix:visual-compare}
\begin{figure*}[ht]
    \centering
    \includegraphics[width=\linewidth]{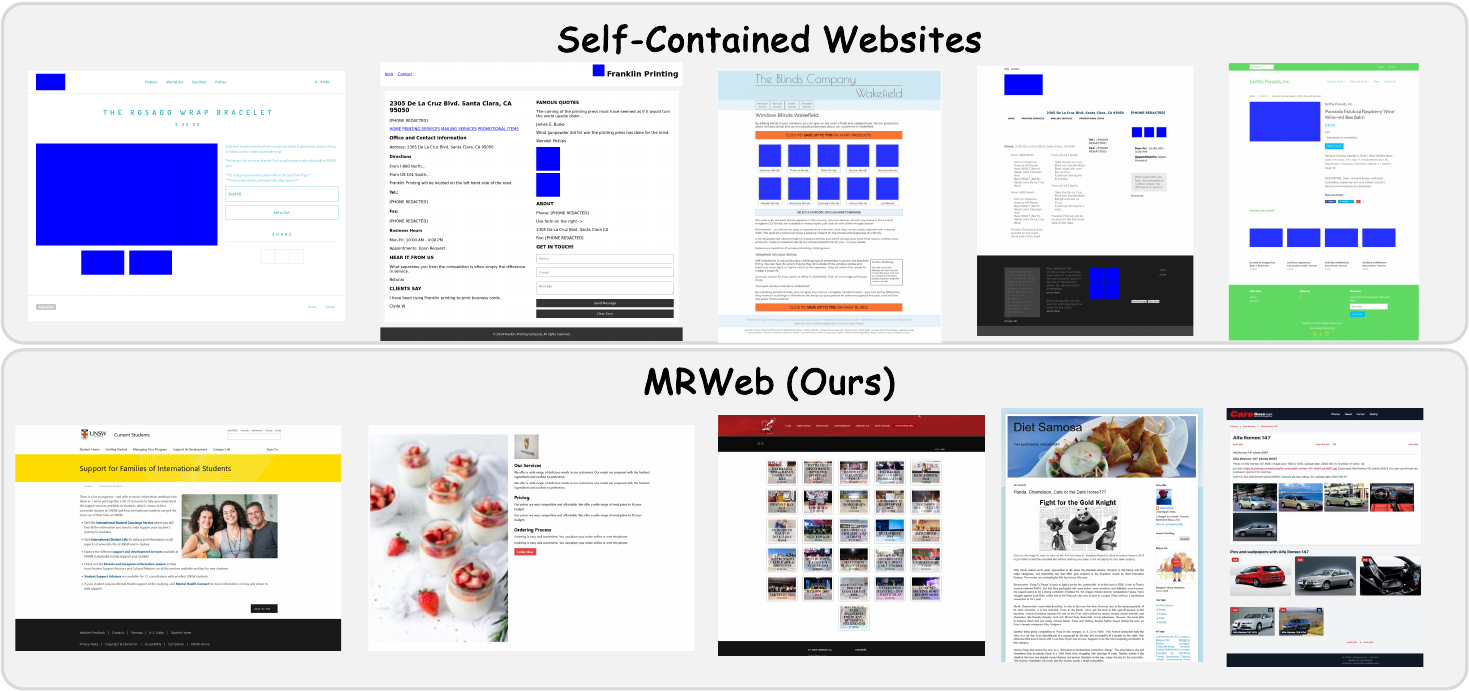}
    \vspace{-5pt}
    \caption{Comparison between self-contained webpages and our multipages resource-aware webpages (MRWebs). Design-to-code webpages contain placeholder images and empty links, whereas MRWebs contain real images and links.}
    \label{fig:data-compare}
    \vspace{-10pt}
\end{figure*}

Figure~\ref{fig:data-compare} shows a comparison between self-contained webpages and our multipages resource-aware webpages (MRWebs). Self-contained webpages contain placeholder images and empty links, whereas MRWebs contain real images and links.

\end{document}